\documentclass[10pt,twocolumn]{article}
\usepackage{amsmath}
\usepackage[sort&compress,square]{natbib}
\setcitestyle{numbers,square}
\usepackage{geometry}
\usepackage{graphicx}
\usepackage{setspace}
\usepackage{flushend}
\begin{document}
\twocolumn[{
\begin{@twocolumnfalse}

\begin{center} 
{\Large \bfseries   Indications of room-temperature superconductivity \\ 
\vspace{0.4 cm}
at a  metal-PZT interface}

\vspace{2cm}
{\mdseries Dhruba Das Gupta}
\linebreak
{\mdseries Department of Physics, University of North Bengal 
\linebreak
Raja Rammohunpur, Siliguri 734013, West Bengal, India}
\vspace{0.5 cm}
\end{center}
\vspace{0.5 cm}
\singlespacing
\begin{center}   
\small
\parbox[t ]{5 in }{  
We report the observation of an exceptionally large room-temperature electrical conductivity in silver and aluminum layers deposited on a lead zirconate titanate (PZT) substrate. The surface resistance of the silver-coated samples also shows a sharp change near 313 K. The results are strongly suggestive of  a   superconductive interfacial layer, and have been interpreted in the framework of Bose-Einstein condensation of bipolarons as the suggested mechanism for  high-temperature superconductivity in cuprates.
}
\end{center}
pacs 74.78.-w, 74.10.+v, 74.81.-g  
\end{@twocolumnfalse}
\vspace{0.9cm}  
}]
\vspace{3cm} 

\section{Introduction}
 
A complete and fully-accepted theory of high-temperature superconductivity, even for the extensively-studied cuprate compounds, is yet to emerge. However, among the different mechanisms suggested for these materials, that of Bose-Einstein condensation of bipolarons has attracted much attention\citep{Mott1,Alexandrov1,Alexandrov2, Alexandrov3}. Its main features, at the simplest level, are the following: 1) Coupling between charge carriers and phonons leading to the formation of polarons and (subsequently) bipolarons in the  $\mathrm{CuO{_2}}$ planes; and 2) Bose-Einstein condensation (BEC) of superlight bipolarons to form the superconducting state. Although some serious objections have been raised against it in the past\cite{Chakraverty1}, the bipolaron model, in a modified form, has found support from a large number of experiments\cite{Muller1, Alexandrov2,Devreese1}.

The early works of Ogg, Schafroth, Butler and Blatt\cite{Ogg1,Ogg2, Schafroth1,Schafroth2, Schafroth3} to explain the phenomenon of superconductivity in terms of BEC were found to be inadequate in explaining the physical properties of the classic superconductors, for which the Bardeen-Cooper-Schrieffer (BCS) theory was highly successful. Discovery of the high-temperature superconductors has brought about a revival of the BEC picture.  

As proposed by Little in his seminal paper\cite{Little1}, a suitable candidate for a superconductor would be a system of long-chain organic molecules with polarizable side chains. This has greatly inspired research in the field of organic superconductors\cite{Jerome1}. A very similar proposal  made  by Ginzburg\cite{Ginzburg1,Ginzburg2} sought to achieve superconductivity in a metallic film, a few atomic layers thick, deposited on a dielectric. According to this idea the electrons in the surface states of the metal, in proximity of the dielectric, would be subject to an attractive interaction by an ``excitonic" mechanism which might lead to superconductivity. For both the above models, the pairing interaction is electronic, in the sense that mediation by phonons is not involved. A large number of experimental results\cite{Keller1} point towards a very important role played by phonons in the cuprates, but some contribution from an electronic mechanism cannot be ruled out\cite{Zhou1,Little2}. 

Lead zirconate titanate (PZT), $\mathrm{Pb(Zr_{\it{x}}Ti_{ {1-\it{x}}})O_3}$, has the perovskite structure and is useful for many applications because of its superior piezo-, pyro- and ferroelectric properties. Its static relative permittivity is of the order of \(10^ {\, 3}\)\cite{Moulson1}.

In this work we report the observation of a region of unusually high conductance in the low-current part of the room-temperature current-voltage characteristics of a thin metallic layer deposited on a PZT substrate. This has been supplemented by an investigation of the surface resistance as a function of temperature. Significance of the results in the light of the above theoretical considerations is discussed.

\section{Experimental Details}

The samples used for current-voltage measurements were (i) thin strips $\approx$ 2 cm $\times$ 2 mm cut off from commercial PZT discs (supplied by Central Electronics Ltd., New Delhi, India) 0.3 mm thick and with an average grain size 1 $\mu$m which were supplied in the poled state and with  $\approx$ 0.1 mm silver coating on both faces and (ii) the same type of strips with the original silver coating removed and $\approx$ 4000A aluminum deposited by vacuum evaporation. The Curie temperature of the material was $360^{0}$C as specified in the manufacturer's data sheet. 

Measurements were carried out at room temperature using a four-probe arrangement with the sample placed inside a double permalloy magnetic shield, the residual magnetic field inside the enclosure being less than $10^{-5}$ tesla. The output voltage, which was of the order of microvolts, was measured using a home-built instrumentation amplifier based on an Analog Devices AD620 chip. Data were recorded in an Agilent 54622A digital storage oscilloscope by using a sawtooth current excitation at a frequency of 20 Hz from a function generator. It was found that scanning near this rate yielded the most consistent and reproducible data, least affected by fluctuations and noise.

Variation of surface resistance as a function of temperature was studied by the single-coil inductance method\cite{Bianconi1,DiCastro1,Gasparov1, Gasparov2}. The sample, which was a 25 mm diameter by 0.3 mm thick PZT disc with silver coating on one side, was placed in close contact (with the silver-coated side facing away) with a planar coil. The coil formed the tank circuit inductance of an RF oscillator whose variation of frequency with temperature was measured. A change in resistivity in any part of the sample would be accompanied by a corresponding change in the reflected impedance from the sample to the coil, and thus produce a change in the oscillator frequency.

\section{Results and Discussion}

Typical current-voltage  characteristics of  the samples (i)  PZT with the original silver coating and (ii) Al-deposited  PZT are shown in Figures 1(a) and 1(b). Data for the actual output as also with a 20-point FFT smoothing have been plotted, the latter being shifted slightly for clarity. It can be seen that up to a current 60 $\mu$A for silver and 0.5 $\mu$A for aluminum films, the voltage across the sample is essentially below the detection threshold. Beyond this range, the voltage increases linearly. The large difference in the threshold currents for the two cases is due to the much higher thickness of the silver coating. Fluctuations in the plot are indicative of the circuit noise and thermal effects associated with the extremely small output voltage ($\approx$ microvolts) and the two-dimensional nature of the system.
 
\begin{figure}[h]
\includegraphics[width=9 cm]{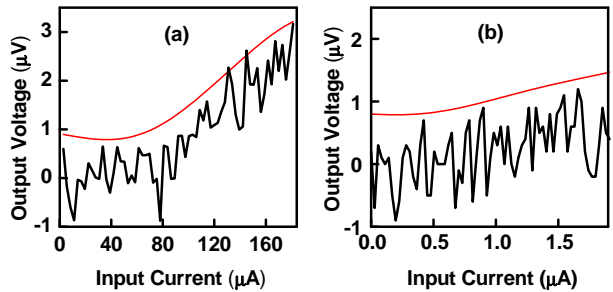} 
\vspace{-6 cm}
\caption{ \selectfont{\small{(Color online)Current-Voltage characteristics of metal films deposited on PZT substrates: (a)Silver (b)Aluminum.}}}
\end{figure}
Figure 2 is a plot showing the variation of oscillator frequency $f$ with temperature $ T $ as discussed above. Increase in $T$ produces a steady decrease in $f$, caused by an increase in the dimensions of the oscillator coil. Superposed on this decrease there is a sharp kink in the $T-f$ curve at $T$$\approx$ 313 K, which  is interpreted  as a rise in resistivity $\rho$ caused by transition to the normal state at the BEC transition temperature  $T_{c}$.
 
\begin{figure}[h]
\includegraphics[width=10 cm]{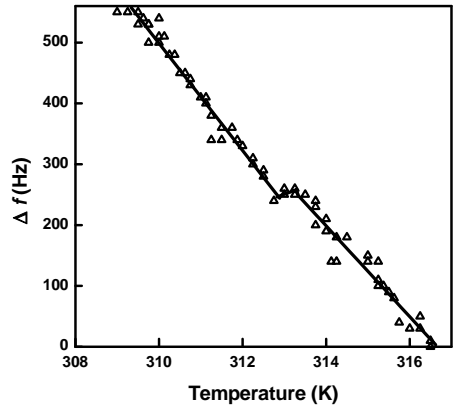}
\vspace{-6 cm} 
\caption{ \selectfont{\small{ Change in oscillator frequency as a function of temperature of silver film on PZT substrate.}}}
\label{multi}
\end{figure}

The above experimental results are consistent with the presence of a superconducting phase within the metal-PZT composite, and it is reasonable to suppose that this phase is located at the interface. We now examine the idea in detail.

The high-temperature cuprate superconductors are all layered compounds and superconductivity arises primarily in the $\mathrm{CuO{_2}}$ planes\cite{Kresin1}. In addition, there is evidence  of formation of charge inhomogeneities such as stripes and checkerboard patterns in these planes\cite{Bianconi2, Tranquada1, Wise1}. The inhomogeneities have been found to be present also in many cases where the sample is non-superconducting\cite{Zheng1, Bianconi3}, and obviously originate from a different mechanism, which we do not discuss here\cite{Mihailovic1}.

We hypothesize that electrons in the metal surface adjoining the PZT disk form an inhomogeneous charge pattern in the same manner as in the high-$T_c$ cuprates. Those in the charge-rich regions form polarons and real-space polaron pairs (bipolarons) by a phononic mechanism, possibly accompanied by an electronic one\cite{Little2, Zhou1}. The large static relative permittivity of PZT is expected to favor the formation of polarons\cite{Devreese1}. The bipolarons, in turn, Bose-condense at a temperature $T_c$ near room temperature. In 2-D, $T_c$ depends on the dimensions of the system, which in this case would be determined by the length scale of the inhomogeneities. Finally, a macroscopic superconducting state is formed by Josephson percolation between the superconducting regions\cite{Mihailovic1}.

A rough estimate of the parameters involved is done as follows. We model a charge-rich region as a square area of linear dimension $l_0 =  2$ nm\cite{Hoffman1} and an effective bipolaron mass $m_b = 10m_e$\cite{Alexandrov4}. For $T_c = 320 K$, the number of bosons, assumed to be non-interacting, comes out as $N_B = 1.8$. The unit cell linear dimension of PZT being roughly 0.5 nm, the area covers 4 $\times$ 4 lattice sites. Thus the number of bosons per lattice site is 0.11, and the condition that there is no overlap between the bosons\cite{Alexandrov3} is clearly satisfied.

The values are purely indicative, and only serve to underline the crucial role of the inhomogeneous charge distribution. At the same time, the exact shape or dimensions of the charge-rich regions play a somewhat secondary role as $T_c$ in 2-D depends only weakly (logarithmically) on the sample area\cite{Ginzburg2}.

\section{Concluding remarks}
That Bose-condensation of bipolarons might lead to superconductivity was first proposed by Alexandrov and Ranninger\cite{Alexandrov5} in 1981. Discovery of the cuprate high-temperature superconductors in 1986 saw the idea being further developed by a number of workers, notably Mott and Alexandrov\cite{Mott3, Alexandrov6,Emin1,Mott4,Mott1, Alexandrov1}. 
  
The experimental results reported here strongly suggest the presence of a superconducting layer near room temperature in the interface between a metal film and a PZT substrate. The data have been interpreted in the framework of the above model in terms of the experimentally-observed inhomogeneous charge patterns in high-temperature superconductors.  

The fact that the results described above have been obtained from very simply-fabricated systems, without the use of any sophisticated set-up and any special attention being given to crystal purity, atomic perfection, lattice matching, etc. suggests that the physical process is a universal one, involving only an interface between a metal and an insulator with a large low-frequency dielectric constant. We note in passing that PZT and the cuprates have similar (perovskite or perovskite-based) crystal structures.   This resemblance may provide an added insight into the basic mechanism of high-temperature superconductivity. 
 
\vspace{1cm}
\bibliographystyle{plainnat}  
 \flushend
\end{document}